# Dual Heuristic Dynamic Programing Control of Grid-Connected Synchronverters

Sepehr Saadatmand, Mohammad Saleh Sanjarinia, Pourya Shamsi, and Mehdi Ferdowsi
Department of Electrical and Computer Engineering
Missouri University of Science and Technology
sszgz@mst.edu, mswvq@mst.edu, shamsip@mst.edu, ferdowsi@mst.edu

*Abstract*—In this paper a new approach to control a grid-connected synchronverter by using a dual heuristic dynamic programing (DHP) design is presented. The disadvantages of conventional synchronverter controller such as the challenges to cope with nonlinearity, uncertainties, and non-inductive grids are discussed. To deal with the aforementioned challenges a neural network–based adaptive critic design is introduced to optimize the associated cost function. The characteristic of the neural networks facilitates the performance under uncertainties and unknown parameters (e.g. different power angles). The proposed DHP design includes three neural networks: system NN, action NN, and critic NN. The simulation results compare the performance of the proposed DHP with a traditional PI-based design and with a neural network predictive controller. It is shown a well-trained DHP design performs in a trajectory, which is more optimal compared to the other two controllers.

*Index Terms*--Dual heuristic dynamic programming, grid-connected inverter, neural network, synchronverter

## I. INTRODUCTION

The energy crisis and the environment pollution challenges have been forcing governments to replace the traditional power generators with renewable energy sources (RESs). The implementation of a three-phase inverter with a dc-link or power storage at the primary side is the most widespread method to connect RESs to the grid. The growing penetration of RESs highlights the importance of three-phase grid-connected inverters. Nowadays, the most common method to govern a grid-connected converter to track the active and reactive power reference is to apply a direct-current control method [1].

Although, the direct-current method is very popular, it has several drawbacks. The fast responding nature of the power electronic inverters and the lack of inertia decreases the power system stability. Moreover, to synchronize with the grid, a phase-luck-loop (PLL) is needed. There have been various studies in introducing new methods to implement PLLs, and several challenges facing uncertainties, noises, and unbalanced conditions have been pointed out [2]. Besides, a direct-current inverter is unable to perform in standalone mode, which is an important characteristic for grid-tied inverters operating in islanded micro-grids. To overcome the aforementioned disadvantages, several solutions have been proposed. Considering the fact that power system inertia is essentially provided by the large kinetic energy buffered in synchronous generators, the concept of virtual synchronous generator (VSG) was introduced recently to virtually imitate the response of the traditional synchronous generators virtually to interact with the power system to improve system inertia, resiliency, microgrid stability, and output impedance [3]-[5]. A VSG-based grid-connected inverter is a nonlinear system. The traditional way to control them is to apply a linear controller for the linearized system model at a specific operating point. By changing the operating point and the system parameters, the controller behavior changes. In other words, a well-designed controller for an operating point might lead to instability for another operating point. Nonlinear controllers (e.g. nonlinear feedback linearization technique) are popular replacements for the linear ones. However, nonlinear controllers are more complicated to design; besides, in some cases like feedback linearization technique, the exact system parameters are needed. Furthermore, if the model of the system varies significantly, it would require an adaptive feed-back linearization scheme which is more challenging than other nonlinear adaptive control techniques. For instance, the application of the neural network-based controllers enables to overcome these liabilities [6], [7].

Neural networks are widely implemented in various areas such as image processing, speech recognition, text mining [8] and control field of study. The artificial neural networks (ANNs) are able to identify a time varying system. With continuous training, the neural network–based model adapts itself through time. The adaptive critic controller (ACD) is a form of reinforcement learning (RL). ACD performs as an optimal controller in an approximate dynamic programing. Dual heuristic programing (DHP) is a powerful critic-based controller [9], [10].

The main contribution of this paper is to present a new method for developing a neural network–based DHP combined with the virtual inertia strategy for controlling a grid-connected inverter. First, a brief review of virtual inertia control concept, frequency control, and the limitations associated with conventional control for VSGs is presented in Section II. The dual heuristic dynamic programming concept and the implementation process in VSGs are explained in Section III. The performance of the proposed DHP controller is evaluated in Section IV. Finally, the conclusion is presented in Section V.

## II. PRINCIPLE AND MODEL OF A VSG

In this section, the VSGs controller structure and the power flow equation are explained. Moreover, it is shown that the virtual inertia controller performance extremely depends on the grid parameters.

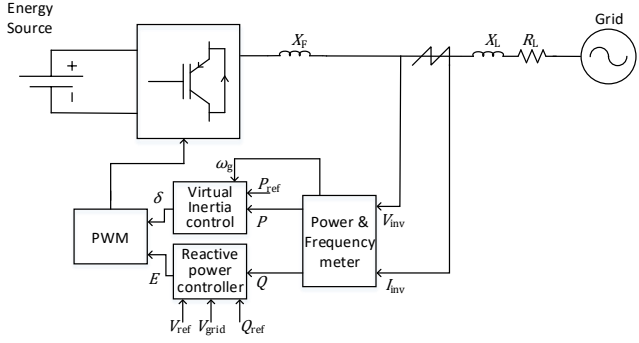

Fig. 1. Conventional Virtual Synchronous Generator (VSG) block diagram

*A. VSG controller*

The control block diagram of the VSG system is illustrated in Fig. 1. In this paper, it is assumed that the inverter is connected to a fixed dc source. Hence, the inverter can route power from the dc source to inject additional power to the grid during transients (not applicable to solar photovoltaic inverters). The control scheme in Fig. 1 is based on the virtual inertia method. In this model, XF, XL, and RL are the inverter output filter reactance, the inverter to the grid line reactance, and the inverter to grid line resistance respectively. The most important component when modelling VSGs is the swing equation of a synchronous generator as follows:

$$P_{in} - P_{out} = J\omega_i \left(\frac{d\omega_i}{dt}\right) + D\Delta\omega \qquad (1)$$

where $P_{in}, P_{out}, \omega_i, J$, and $D$ are the input power to VSG, the electric output power, the angular velocity of the virtual rotor, the virtual rotor moment of inertia, and the droop coefficient, respectively. In this equation, $\Delta\omega$ is defined by $\Delta\omega = \omega_i - \omega_g$ where $\omega_g$ is the grid angular velocity while the inverter is connected to the grid or the reference angular velocity while the inverter works in a standalone mode.

The command signal to the inverter includes two parts. First, it needs the RMS value of the inverter voltage or peak value of inverter phase voltage ($E$). Secondly, it needs the inverter power angle with respect to the grid ($\delta$). In order to compute E, the electrical output power can be computed by measuring the inverter voltage signals and the current signal injected into the grid. Having all the necessary parameters in (1), $\Delta\omega$ can be computed at each control cycle. Then, the mechanical phase can be calculated by integrating this frequency.

$$\delta = \int \Delta\omega \cdot dt$$

In high voltage power systems, in order to tune the inverter voltage, a reactive power controller with a voltage droop is utilized. Incorporation of a voltage droop and an integrator controller provides the RMS/peak value of the voltage using the following equation:

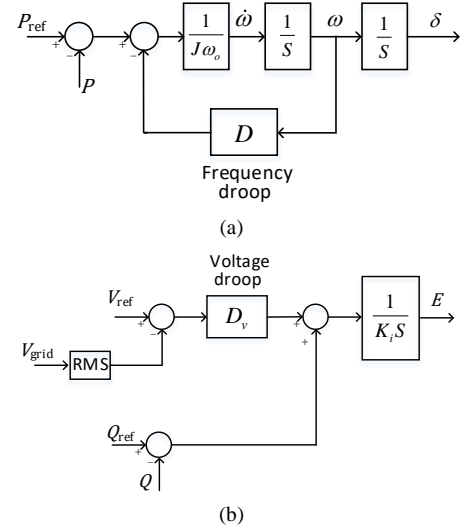

Fig. 2. VSG controllers. (a) Power-frequency control block diagram. (b) Voltage-reactive power control block diagram

$$E = \frac{1}{K_i} \int \Delta Q \cdot dt - D_v \Delta V$$

where $K_i$ and $D_v$ are the integrator coefficient and the voltage droop, respectively. The inverter reactive power tracking error is given by $\Delta Q = Q_{ref} - Q_e$, and the inverter voltage tracking error is given by $\Delta V = V_{ref} - V_i$. The reference reactive power for the inverter is set to $Q_{ref}$ and the inverter output reactive power can be computed by a power meter block. The variable $V_i$ is the inverter output voltage, and $V_{ref}$ is the reference voltage for the inverter. Fig. 2 shows the active and reactive power controller block diagrams.

*B. Power flow equation for grid-connected inverters*

The proposed VSG average model can be derived based on a voltage source as shown in Fig. 3. In the figure, $X_{eq}$ is the equivalent reactance per phase (line and filter) and can be computed as $X_{eq} = X_F + X_L$, $R_{eq}$ presents the equivalent resistance per phase (line and filter) given as $R_{eq} = R_L$, and $Z_{eq}$ is the equivalent impedance per phase (line and filter) given as $Z_{eq} = jX_{eq} + R_{eq}$. The active and reactive power delivered by the converter to the grid can be expressed as

$$P = \frac{1}{2}\left[\left(\frac{E^2}{Z_{eq}^2} - \frac{EV\cos\delta}{Z_{eq}^2}\right)R_{eq} + \frac{EV}{Z_{eq}^2}X_{eq}\sin\delta\right]$$
$$Q = \frac{1}{2}\left[\left(\frac{E^2}{Z_{eq}^2} - \frac{EV\cos\delta}{Z_{eq}^2}\right)X_{eq} - \frac{EV}{Z_{eq}^2}R_{eq}\sin\delta\right]$$

where $P$ and $Q$ are the delivered active and reactive power (per phas), V is the peak value of the phase voltage of the grid, E is the peak value of the output voltage of the inverter and $\delta$ is the phase angle between the grid voltage and the inverter voltage. For an inductive equivalent impedance (i.e. $X_{eq} \gg R_{eq}$ as in [11], [12],) the active and reactive power can be estimated as:

$$P \approx \frac{EV}{2X_{eq}}\sin\delta \qquad (2)$$

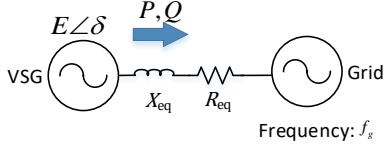

Fig. 3. Equivalent circuit diagram of a grid connected VSG

$$Q \approx \frac{E}{2X_{eq}}(E - V\cos\delta) \quad (3)$$

Generally, the inverter power angle $\delta$ is small, and $\sin\delta$ can be approximated by $\delta$, and $\cos\delta$ can be approximated by 1. Therefore, (2) and (3) can be written as

$$P \approx \frac{EV}{2X_{eq}}\delta \quad (4)$$

$$Q \approx \frac{E}{2X_{eq}}(E - V) \quad (5)$$

Equation (4) and (5) clarify that in inductive grids, the active power is proportional to the power angle and the reactive power is proportional to the inverter voltage. In this case, the conventional VSG controller performance is acceptable; nonetheless, in low-voltage grids that are mostly resistive or semi-resistive, this assumption is no longer valid. In other words, Q depends on both the power angle and the voltage magnitude. In order to turn the reactive power controller for non-inductive grids, all the parameters of the system model need to be known to make it possible to design an acceptable reactive power controller. However, in the power system, the inverter might face uncertainties such as line impedance changes or nonlinear behaviors (e.g. transformer saturation) in electrical element that alter the reactive power equation. In this paper, an adaptive dynamic controller, capable of adjusting its parameter, is used to find the optimal solution and the results are compared with the conventional controller performance.

### III. DUAL HEURISTIC DYNAMIC PROGRAMING

ACDs are neural network–based tools for optimization over time. The highlighted feature of ACDs is their ability to perform under conditions of noise and uncertainties. A family of ACDs by combining the reinforcement learning and dynamic programing is proposed in [13]. A typical ACD consists of two subnetworks: action network and critic network. The action network and the critic network can be connected together through an identification model (model-dependent design) or directly (action-dependent design). Fig. 4 depicts the block diagram of a simple ACD. The action network objective is to generate a set of control to optimize a utility function over time, and the critic network objective is to criticize how good the action network performs. There are four main classes for implementing ACDs known as heuristic dynamic programing (HDP), dual heuristic dynamic programing (DHP), global heuristic dynamic programing (GHDP), and global dual heuristic dynamic programing (GDHP). This paper presents the DHP model and compares its performance with the traditional PI and neural network predictive controller (NNPC).

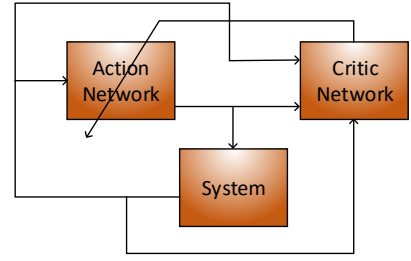

Fig. 4. Adaptive critic design block diagram

Assuming that the optimal policy can be defined as a differentiable function of the state variables, dynamic programing provides a set of control or a control policy to minimize the cost-to-go function defined as

$$J(t) = \sum_{k=0}^{\infty} \gamma^i U(t+k) \quad (6)$$

where $\gamma$ is a discount factor, ($0 < \gamma < 1$), to make sure that the cost-to-go is bounded and $U(\cdot)$ is the utility function. By rewriting (6) in the form of Bellman's Recursion, it can be presented as follows:

$$J(k) = U(k) + \gamma J(k+1) \quad (7)$$

In ACDs, the main objective of the critic network is to feed the derivative of the cost-to-go function with respect to the control signal to the action network.

The utility function in this paper is defined as

$$U(k) = \sqrt{K_P e_P^2 + K_Q e_Q^2 + K_f e_f^2}$$

where $e_P, e_Q, e_f$ are the error signals for active power, the reactive power, and the frequency, respectively. These error can be defined as

$$e_P = P_{set} - P$$
$$e_Q = Q_{set} - Q$$
$$e_f = f_g - f$$

and $K_P, K_Q, K_f$ are the active power coefficient, the reactive power coefficient, and the frequency coefficient, respectively. These coefficients can also be defined as the weight matrix in a normalized function to define the importance of each error signal. Mathematically solving the dynamic programing is complex and expensive. ACDs proposed a technique to provide the optimal control set to minimize $J(\cdot)$. In order to train the neural network, the derivative of the error or cost-to-go function is needed to criticize how well the action network is functioning. For example, the critic network in HDP method estimates the cost function and then by taking its derivative, the feedback signal to the action network is generated.

Fig. 5 shows the block diagram of a critic-based DHP controller. In this figure, $X(t)$ is the state vector, and $u(t)$ is the control vector, which is generated by the action neural network. By feeding the action network control signal to the system/plant the next state vector can be measured/computed as $X(t + Ts)$, where $Ts$ is the sampling time. The main objective of the critic network is to provide the necessary feedback to the action network to make sure the control outputs satisfy the plant control objective. As mentioned, the main objective of the designed controller is to minimize $J(t)$, and in order to do that the critic neural network provides the action network with the gradient of $J(t)$ with respect to the state vector, shown with letter $\lambda$. In DHPs, the critic network estimates the cost-to-go derivatives with respect to the states directly.

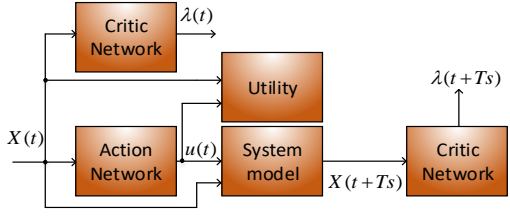

Fig. 5. Block diagram of the proposed controller for VSG

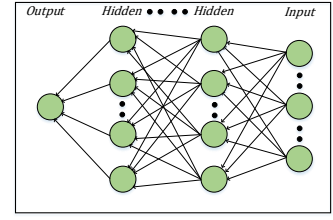

Fig. 6. Fully connected feedforward neural network

### A. Critic neural network

As mentioned, the critic network objective is to estimate the gradient of cost-to-go function with respect to the system states. By taking the derivative of (7) as

$$\frac{\partial}{\partial X_i(t)} J(t) = \frac{\partial}{\partial X_i(t)} (U(t) + \gamma J(t+1)). \tag{8}$$

In order to train the critic network the error expressed as

$$\|Er\| = \sum_t e_c^T(t) e_c(t), \tag{9}$$

needs to be minimized over time period t. In (9), $e_c$ at each period is defined as follows:

$$e_c(t) = \frac{\partial}{\partial X(t)} J(t) - \frac{\partial}{\partial X(t)} (U(t) + \gamma J(t+1)). \tag{10}$$

In addition, by applying the chain rule in DHP, (10) can be written as follows:

$$\frac{\partial J(t+1)}{\partial X_j(t)} = \sum_{i=1}^{n} \lambda_i(t+1) \frac{\partial X_i(t+1)}{\partial X_j(t)} + \sum_{k=1}^{m} \sum_{i=1}^{n} \lambda_i(t+1) \frac{\partial X_i(t+1)}{\partial u_k(t)} \frac{\partial u_k(t)}{\partial X_j(t)} \tag{11}$$

where n is the number of states, m is the number of controls, and $\lambda_i(t+1) = \partial J(t+1)/(\partial X_i(t+1))$. In this paper, state vector is defined as $X = [P\ Q\ e_p\ e_q\ e_f\ \theta_i]$ and the control signal is the inverter voltage magnitude. By implementing (11) in (10), it can be written as

$$e_{c_j}(t) = \frac{\partial J(t)}{\partial X_j(t)} - \frac{\partial}{\partial X(t)} (U(t) + \gamma J(t+1)). \tag{12}$$

Equation (12) is used to train the critic network. In order to evaluate the right-hand side of (12), the exact system model is needed to compute the partial derivative of next state with respect to the current state. To do so, there are two solutions. First, if the system model in known and all the parameters are certain, the derivative can be directly computed. Second, if the system parameters are not certain, a neural network can be used as a system identifier. By training the system neural network, the aforementioned derivative can be computed and used. In this paper, it is assumed that the parameters are not certain and a pretrained fully connected forward neural network with two hidden layers consisting of five nodes at each layer is used to model the system. Fig. 7 illustrates a fully connected forward neural network.

### B. Action neural network

The goal of the action neural network is to generate the control signal to minimize the cost-to-go function for the immediate future. In other words, the objective is to minimize the sum of the utility function over a period of 1000s in this paper. The action neural network implementation is similar to that of critic neural network. A fully connected multi-layer feedforward neural network with two hidden layers, each with eight nodes, and one output is used to create the action network. The input signal to this network is the state vector. The output signal of the action NN is the peak value of the output voltage of the inverter. In order to update weights in the action NN, the backpropagation algorithm is used. The objective is to minimize $J(k)$ as follows:

$$\zeta = \sum_k \frac{\partial J(k+1)}{\partial u(k)}.$$

The gradient of $J(\cdot)$ is given by the critic network. This gradient is used to updates the weights of action neural network.

## IV. PERFORMANCE EVALUATION OF THE TRAINED DHP VIRTUAL INERTIA–BASED CONTROLLER

The block diagram of DHP-based synchronverter is shown in Fig. 7. As shown, a neural network–based DHP is implemented to generate the voltage magnitude of the inverter. As mentioned, there are three neural networks in this controller, which need to be trained. The system neural network needs to be pretrained and to do so random references are generated as inputs to the system while a PI controls the synchronverter. The objective of the system neural network is to estimate the system model to extract the aforementioned derivatives. A set of data with the length of 10000 is used to train the system neural network. The gradient descent method used with the backpropagation to update the weights and train the network.

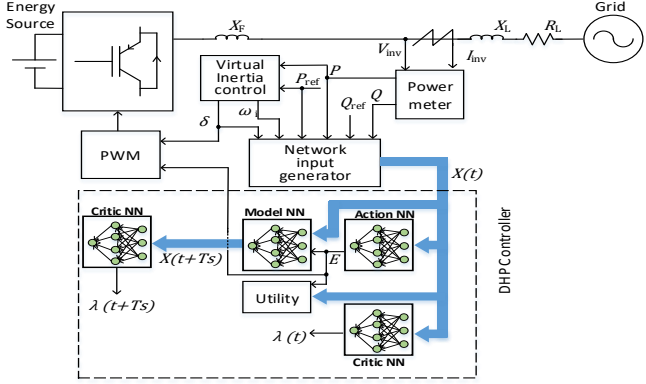

Fig. 7. DHP-based synchronverter block diagram

The importance of the ability to work in both inductive and resistive grids is discussed in Section II. The controller is trained and tested for both inductive and resistive grids and the results are shown in this section. The parameters that are used for the simulation are shown in Table I. In this paper, DHP design, PI controller, and NNPC are implemented to compare their results. NNPC is a supervised control method. NNPC goal is to optimize the same cost function over the same time horizon. A neural network is used to identify the system model, similar to DHP, and by having ten different choices to change the current voltage, the best voltage option is chosen to minimize the cost function.

### A. Inductive grid

As it is discussed and expected the PI-based synchronverter performs relatively well in tracking the power references. However, as shown in Fig. 8, the results for both DHP and NNPC controller are better than that of PI. Moreover, DHP-based synchronverter performance is slightly better than NNPC because, the DHP base search the entire acceptable domain for the control signal while the NNPC select the result from a set with only ten choices. In this paper, four different operating points, active power and reactive power references, are generated to analyze the performance of each controller. In Fig. 8 (a) and (b), the active and the reactive power tracking ability of each controller are illustrated. In addition, Fig. 8c, 8d, 8e show the magnified version of the active power, the reactive power, and the frequency response, respectively.

TABLE I
SYSTEM PARAMETERS USED IN SIMULATION

| Parameter | Value | Unit |
|---|---|---|
| DC voltage | 250 | V |
| AC line voltage | 110 | V |
| AC frequency | 60 | Hz |
| Moment of inertia | 0.1 | Kg.m2 |
| Frequency droop | %4 | -- |
| Inverter power rating | 5 | kW |
| Inductive line | | |
| Filter inductance | 1 | µH |
| Line inductance | 100 | µH |
| Line resistance | 10 | mΩ |
| Resistive line | | |
| Filter inductance | 1 | µH |
| Line inductance | 1 | µH |
| Line resistance | 500 | mΩ |
| DHP parameters | | |
| γ | 1 | |
| Sampling time | 1 | ms |
| [$k_P$ $k_Q$ $k_f$] | [1 1 0] | -- |

### A. Resistive grid

As discussed, the performance of PI-based synchronverter in resistive grid grids is challenging. The reason is that, the reactive power and the active power control are not completely decoupled. The neural network nature of the DHP and NNPC helps to overcome this problem. As expected, the performance of DHP and NNPC are exceptionally better than a PI-based. Fig. 9 illustrates the performance of the mentioned three controllers. In this figure, (a) and (b) illustrate the performance of the mentioned controllers in tracking the active and the reactive power references, respectively. To be able to compare the results better, the magnified version of the active and the reactive tracking is shown in part (c) and (d), respectively. As shown, the performance of DHP and NNPC is tremendously better than PI-base especially in tracking the reactive power. As discussed earlier, the DHP performance is slightly better than the NNPC thanks to its ability to search the entire domain for the optimal control signal.

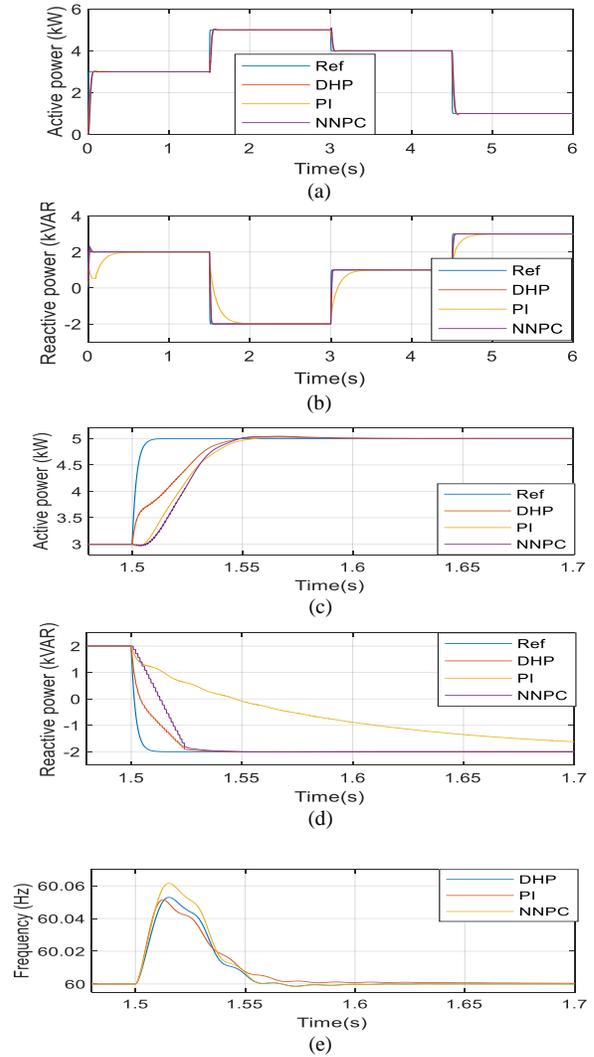

Fig. 8. DHP,PI, and NNPC designs performance for a VSG connected to the inductive grid (a) active power (b) reactive power (c) active power (zoomed in) (d) reactive power (zoomed in) (e) frequency response

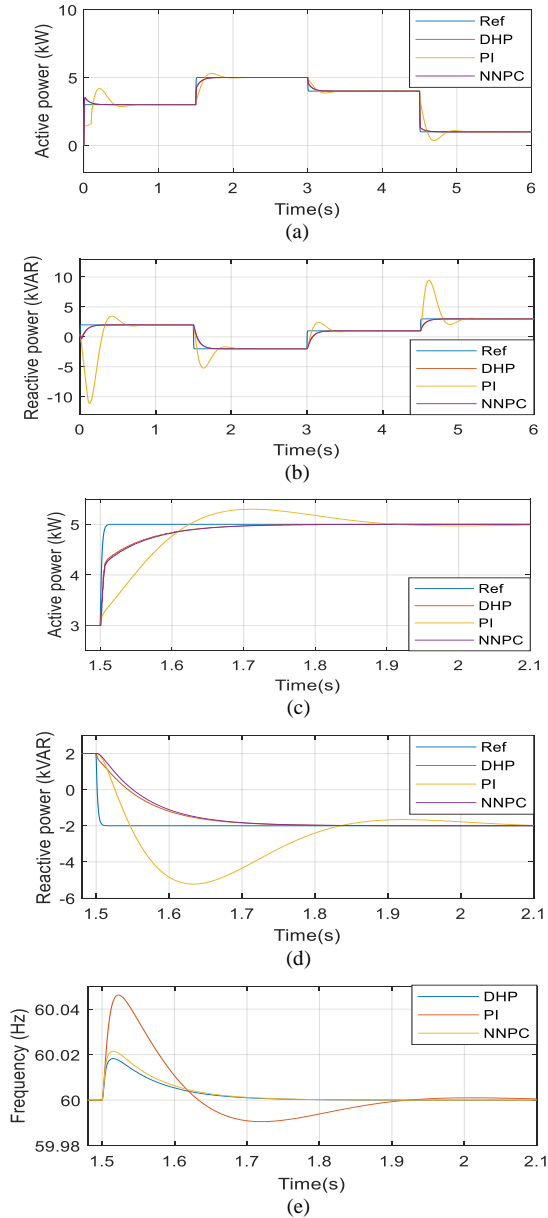

Fig. 9. DHP, PI, and NNPC designs performance for a VSG connected to the resistive grid (a) active power (b) reactive power (c) active power (zoomed in) (d) reactive power (zoomed in) (e) frequency response

## V. CONCLUSION

Controlling a grid-connected inverter has been studied for years and various approaches have been presented. Virtual synchronous generator approach has been introduced to overcome the disadvantages of the traditional inverter controllers. The conventional PI-based design to control the voltage magnitude of the inverter is unable to perform under uncertainties, nonlinear systems, and especially in resistive grids. Dual heuristic dynamic programing is a neural network–based approximate dynamic programing technique that can replace the conventional PIs. In this paper, a DHP control scheme was utilized to control a synchroinverter. In order to operate on both inductive and resistive grids, active power was also added as an input to the proposed controller. The simulation results show the tremendous improvement in tracking ability of the inverter by replacing a PI controller with a neural network–based one, especially DHP designs.

## II. ACKNOWLEDGMENT

This material is based upon work supported by the U.S. Department of Energy, "Enabling Extreme Fast Charging with Energy Storage", DE-EE0008449.